# Low Distortion Fusion Bonding using Pneumatically Warped Wafers

Utkarsh Jain
*imec*
Leuven, Belgium
utkarsh.jain@imec.be

Koen D'havé
*imec*
Leuven, Belgium
koen.dhave@imec.be

Philipp Schmidt
*SUSS*
Sternenfels, Germany
philipp.schmidt@suss.com

Damien Leech
*imec*
Leuven, Belgium
damien.leech@imec.be

Serena Iacovo
*imec*
Leuven, Belgium
serena.iacovo@imec.be

Philippe Muller
*SUSS*
Sternenfels, Germany
philippe.muller@suss.com

Dennis Bumuller
*SUSS*
Sternenfels, Germany
dennis.bumuller@suss.com

Thomas Schmidt
*SUSS*
Sternenfels, Germany
thomas.schmidt@suss.com

Koen Kennes
*imec*
Leuven, Belgium
koen.kennes@imec.be

Steven Brems
*imec*
Leuven, Belgium
steven.brems@imec.be

Eric Beyne
*imec*
Leuven, Belgium
eric.beyne@imec.be

***Abstract*— Backside power-delivery-network (BSPDN) schemes require wafer-to-wafer bonding steps that do not leave high order shape changes or localized stresses in the bonded stack to ensure that these don't get transferred to the thinned wafer for further lithographic exposure. However, bonding mechanics involve strong adhesive forces which can inherently create localized distortions that lithography tools must eventually compensate. Some contributors to grid distortion of the target wafer are, method of bond initiation, bond front velocity variations, and lack of symmetry between the wafers. In this work, we evaluate a low-distortion bonding approach in the SUSS XBA tool, where some of these contributors are tackled at the source, namely by initiating the bond without a localized external force, and keeping the wafers compliant & symmetric during bonding. Wafers are bonded with a slight pre-stress due to controlled gas pressure applied over the whole backside of the wafers throughout bonding. Wafer-shape measurements of the bonded stacks are used to perform gradient-based in-plane-displacements (IPD) modelling to estimate bonding-induced grid distortion. We identify a persistent ring-shaped distortion region on the wafer - whose origin we suspect to originate from bonding mechanics itself – the bondwave's interruption. Dense scanner metrology is used on patterned-bonded wafers to confirm the location and severity of distortion predictions from patterned wafer geometry (PWG) measurements. On the PWG distortion maps and scanner grid readouts, we perform alignment and CPE modelling with different field layouts. Sub-10 nm levels of residual grid distortion are achievable with relatively low-order correction models, and ≤ 3 nm by using advanced CPE models. These results demonstrate that pneumatically warped bonding yields low-distortion bonded stacks, and that simple process tuning can decouple the dominant distortion source from edge-related variability.**

***Keywords*— *bonding induced distortion, wafer bonding, fusion bonding, direct bonding, back side lithography, BS-PDN***

## I. Introduction

Wafer-to-wafer bonding is becoming an essential building block for advanced logic integration, particularly in schemes that rely on post-bond thinning and backside patterning [1,2]. One of the most promising pathways extending the logic roadmap is the backside power-delivery-network (BSPDN) concept at the standard-cell level, where a lithography step is performed on the thinned backside of a bonded pair. This application requires stringent overlay requirements between features defined before and after bonding [3].

Meeting these alignment targets is complicated by the fact that wafer bonding inherently modifies wafer geometry, creating a fundamental bottleneck for BSPDN implementation. The act of pressing two wafers into contact—together with their pre-existing shapes, stress states, and bonding dynamics—introduces distortions that persist in the stack and alter the lithography target map [4]. While scanners can correct low-order components of the resulting grid deformation, the higher-order components require dense metrology to quantify and are therefore more difficult to anticipate or compensate. These higher-order components arise from several physical contributors, including the method of bond initiation, variations in bond-front velocity, hardware imprints, or any other localized stress residues in the bonded stack which can even arise from bonder-chuck contamination.

Given the growing relevance of backside patterning and the need for predictable overlay behavior after thinning, understanding how bonding mechanics shape the distortion map is essential. In particular, bonding configurations that minimize high-order shape changes are of interest for enabling reproducible, low-distortion bonded stacks which would shift distribution of backside (BS) print overlay content to lower orders of the correction content, which could minimize the use of dense metrology on the BS.

In this work, we investigate such a bonding approach and evaluate its impact on wafer-scale distortion using complementary metrology techniques. To quantify distortion, we combine:

1. double-thickness wafer-shape characterization using PWG interferometry, followed by gradient-based (1$^{st}$

order derivatives) analysis to infer the (would-be chucking induced) in-plane-displacements; and curvature-based (2nd order derivatives) stress fields;

2. direct grid readouts on patterned wafers after thinning, enabling scanner metrology which ultimately determines the target grid for exposure.

The paper characterizes the resulting distortion signatures, identifies the physical origins of localized deformation, and examines the extent to which scanner alignment and correction models can compensate these effects in the context of BSPDN-type applications. We apply standard 6-parameter wafer-alignment models to separate low-order distortion components from the residuals. Most distortion introduced during bonding appears as low-order terms, which scanners can readily calibrate and correct. The remaining higher-order distortion is symmetrically distributed (over the wafer scale) but localized at some radial locations. These localized regions can be further reduced using scanner correction-per-exposure (CPE) models. Different field layouts have been used to indicate the suitability of this bonding technique for, for example use advanced packaging lithography, or advanced lithography in immersion or EUV scanner field layouts for advanced logic nodes.

## II. Bonds using unpatterned wafers with SiCN dielectric finish

### A. *Wafer bonding surface preparation*

Both the blanket and patterned (see later section) wafers' bonding surface were prepared with PECVD SiCN deposition of approximately 120 nm thickness. The wafers then underwent post deposition annealing for 120 min at 350 °C in $N_2$ atmosphere. Finally they were subjected to CMP, which reduces the effective thickness to approximately 90 nm, and yields ultra smooth roughness ~0.1 nm.

Prior to bonding, each wafer is rinsed with DI water, and undergoes surface activation in a dual frequency low pressure $N_2$ plasma chamber in the XBS300 cluster from SUSS at 100 W. Then they again undergo DI water rinsing and are spin-dried until a minimal amount of water molecules are left adsorbed on the surface to enable hydrophilic bonding.

### B. *Wafer chucking, bond initiation and bond front progression*

The wafers are carried in the XBS300 cluster with a robot capable of handing both single and double-thickness wafers. In this case, the single wafers are carried between the different cleaner and plasma chambers as described in previous subsection, and before being carried into the bond chamber (XBA). The blanket wafers are loaded after pre-alignment station into the bond. While the patterned wafers (discussed later) can be marker-aligned in the XBA prior to bond initiation. The wafers are vacuum chucked along a circular rim only close to the edges. A small gas pressure (called '*contact pressure*', few millibars) is applied from the center of the chucks towards the wafer backside such that they bulge outwards away from the chuck, as shown Fig 1. A higher pressure results in a higher warp at the center, which can help ensure that the first contact between the wafers happens at the center (assuming no lack of symmetry from the stresses in the wafers and the chucks' hold on the wafers). The magnitude of contact pressure applied also determines how compliant the wafers remain during bonding – resulting in the bond front's propagation resembling the ideal case of free-hanging wafers.

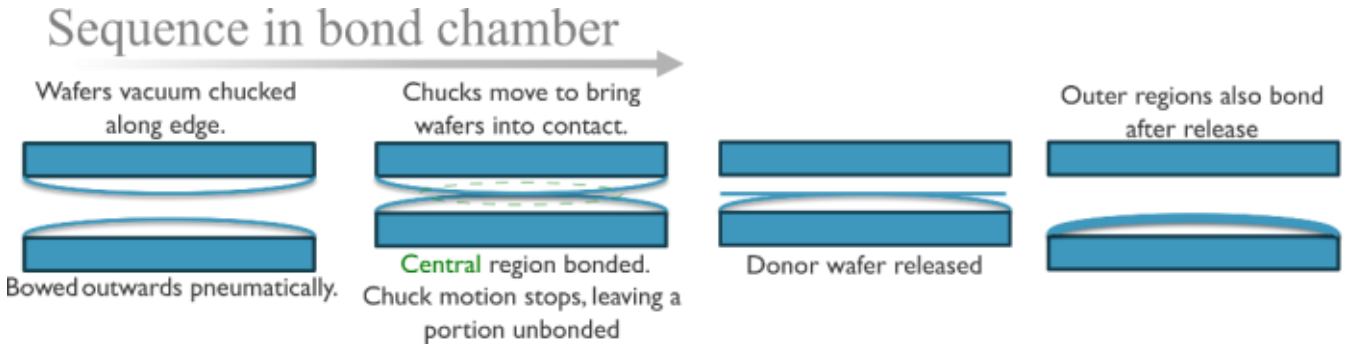

*Fig. 1* Schematic of the bonding sequence once wafers prepared with the appropriate surface treatments are chucked in the bond chamber. Wafers are bulged towards each other by a small gas pressure coming from the backside. The chucks move towards each other to initiate the bond.

The bond is initiated by moving the chucks towards each other with a prescribed velocity (10s of µm/s). Since the work of adhesion of these atomically smooth surfaces prepared with CMP, plasma activation and water rinse is very high (~0.1 J/m$^2$, [5]), the wafers instantly adhere and a bond front spontaneously propagates as far as the chucking conditions allow. The chucks can be programmed to stop moving towards each other such that the wafers are not compressed together over their whole surface. This is implemented in XBA's control software by programming the '*final gap'* (henceforth called parameter $G$) between the unbonded parts of the wafer (close to the edge) which the bond chamber is allowed to reach – typically up to 100 $\mu$m. Once this stage is reached, the vacuum on one or both the chucks can be released so the bond front can propagate to the wafers' edge, completing the process.

## III. Characterisation of blanket bonded wafers

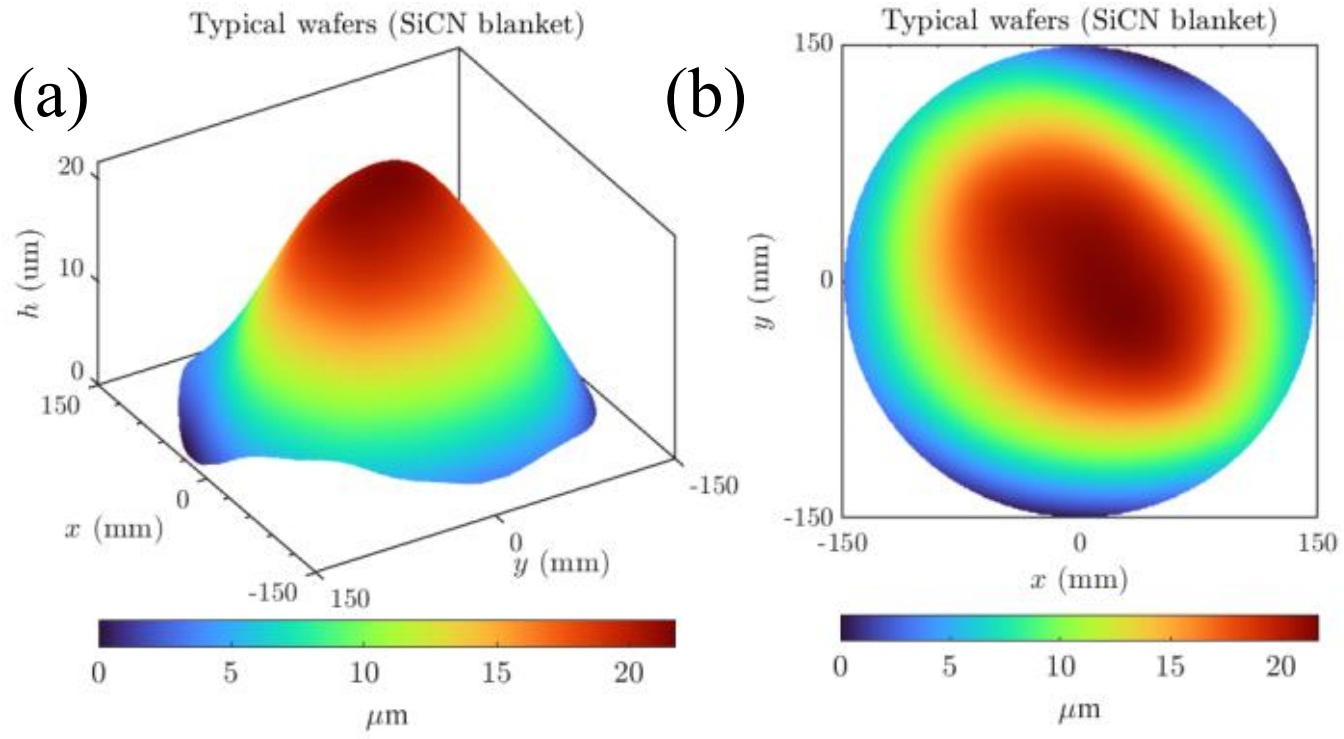

*Fig. 2* Typical shapes of the blanket SiCN wafers used for bonding tests. Shapes measured in a PWG tool. Panel (a) shows the 3D view while panel (b) shows planar view of the same wafer. After deposition of the bonding interface on bare Si, warps of upto ~20 um result. In all wafer plots here and henceforth, data outside 147 mm radius has been discarded.

Wafer shapes both pre- and post-bonding were measured in KLA's PWG5, a dual-side Fizeau interferometry-based tool. Both sides of the wafers can be simultaneously measured in this tool, while any gravity-induced warping is neutralized by holding the wafers vertically. Typical shapes of unbonded SiCN blanket wafers are shown in Fig 2. The surface preparation (PECVD, post-deposition anneal, and CMP) create a slight umbrella shape of the wafers of ~20$\mu$m peak-to-peak. When these wafers are chucked in XBA, the contact pressure bends them by a few hundred microns. Thus, the effects of small warpage variation in incoming wafers are suppressed by the backside pressure prior to bonding initiation in XBA. By tuning the contact pressures on the two chucks, relatively flat bonded wafers can be produced by suppressing the kind of incoming warps as shown in Fig 2.

### *A. Bonded wafers shapes using PWG*

PWG5 measurements were also done on all wafers post-bonding. Typical bonded wafer shape (thickness 2×775 $\mu$m) is shown in Fig 3(a), both in 3D and a 2D plane view. Some asymmetries persist between the wafers during bonding initiation despite achieving a relatively flat 6 $\mu$m warped bonded wafer.

In the raw wafer shape (Fig 3(a)) a nearly symmetric, faint ring shape is visible. The raw shape can be post-processed using inbuilt algorithms in PWG5 to extract nanotopography of the back (or front) surfaces of the wafer. In this case, the back side of the bonded wafer is shown in Fig 3(b). The principle behind determining the nanotopography on the surface is first applying a low-pass Gaussian filter to the raw shape data in both x-y directions and subtracting it from the raw shape. The nanotopography map reveals a smooth surface spanning much of the wafer. However the faint ring shape present in the raw shape (Fig 3(a)) is clearly highlighted as a step change in nanotopography.

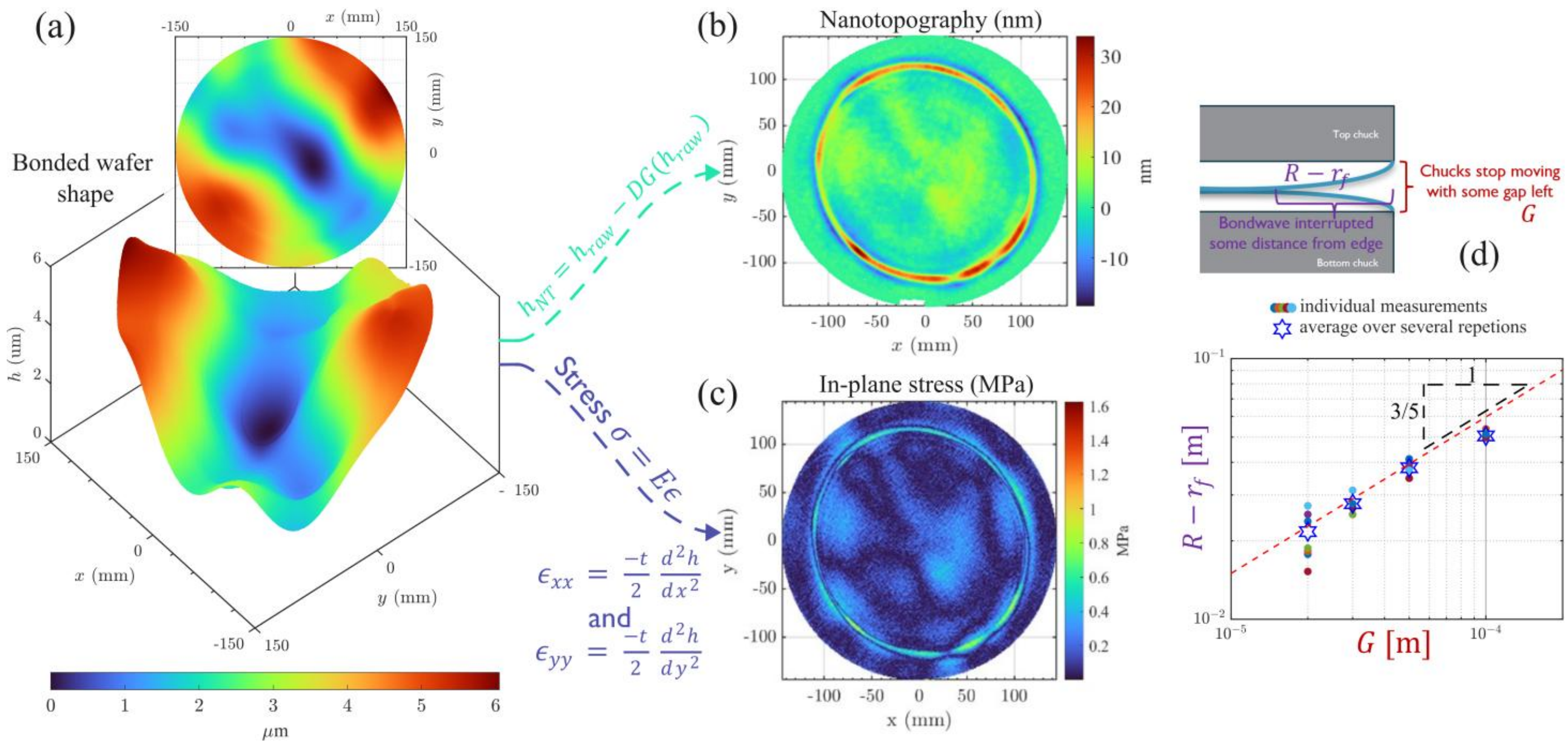

*Fig. 3* (a) Typical shape of bonded wafers in this bond chamber. Relatively flat bonded (double-thickness, 2×775 $\mu$m) wafers can be achieved by tuning the background pneumatic pressures when the wafers are chucked in the tool prior to bond initiation. The main figure shows the 3D shape while the inset shows the same map in (*x,y*) plane. A ring shape is visible in parts of panel (a) at some distance from the edge. (b)Nanotopography map of the bonded wafer is determined by first applying a low pass filter (double-Gaussian filter, 20mm wide cut-off) to the raw wafer shape, $DG(h_{raw})$ and subtracting it from the raw wafer shape $h_{raw}$. A distinct step change of ~10s of nm in the bonded-stack is revealed. (c)The raw wafer shape is subjected to double derivatives in (*x,y*) coordinates, and using plane-strain theory, the in-plane strain ($\epsilon_{xx}, \epsilon_{yy}$) in the stack can be approximated. Strains ($\epsilon_{xx}, \epsilon_{yy}$) can be multiplied by elastic modulus of the substrates (~130 GPa) to plot the approximate in-plane stress in the bonded stack. A value of ~1 MPa suggests an excess stress built by unbalanced adhesive forces near the bond front, due to interruption of the bondwave propagation. (d)The relation between remaining gap where the bonder chucks stop moving ($G$, 20—100 $\mu$m, prior to vacuum release) and the ring fingerprint's location ($R - r_f$, 1—4 cm from the edge) is plotted, where $R$ is the wafer radius (0.15 m). Stopping the chucks' movement towards each other stops the bondwave farther from the wafer edge. It approximately follows the power law 3/5.

### *B. In-plane strain and stress*

This ring-shaped region is further investigated by calculating the in-plane stress in the double-thickness stack by first calculating the in-plane strains in the stack using plane-strain theory, which consists of computing local second-order derivatives:

$$\epsilon_{xx} = \frac{-t}{2}\,\frac{d^2h}{dx^2} \text{ and } \epsilon_{yy} = \frac{-t}{2}\,\frac{d^2h}{dy^2}, \tag{1}$$

where $t$ is the thickness of the concerned stack (2×775 $\mu$m). Stress is simply determined as $\sigma = \epsilon \times E$, where $E$ is the Young's modulus of the material (130 GPa). The stress map in Fig 3(c) shows ~MPa of residual stress distributed roughly axisymmetrically in the stack. The near-axisymmetric distribution is suggestive of this fingerprint being a residue of bond front propagation itself. While the magnitude, which is roughly the magnitude of weak van der Waals' forces [6], is

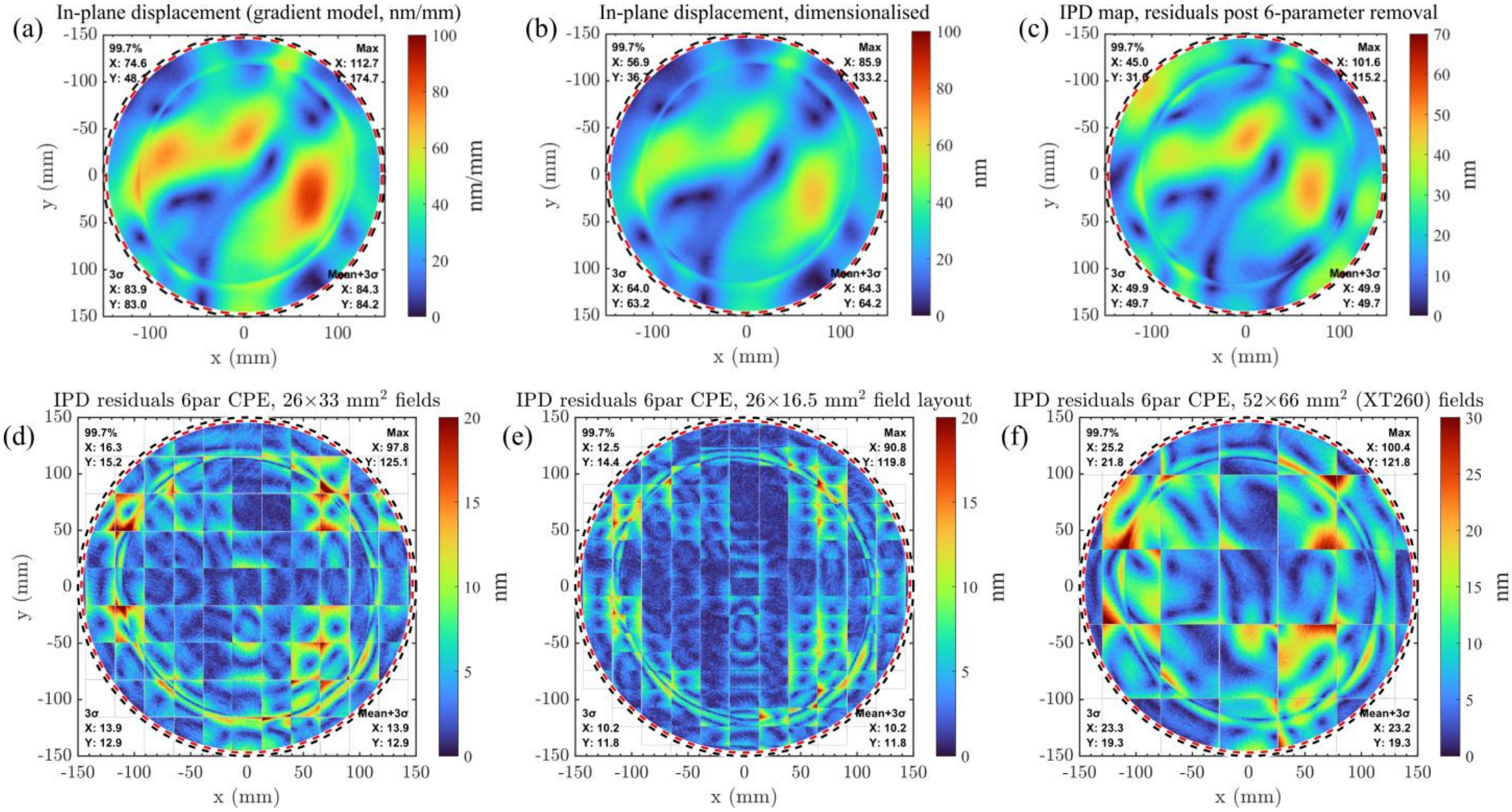

*Fig. 4* (a)First-derivates of the shape of the bonded wafer (double thickness) can be used to infer 'in-plane-displacement' (IPD) maps as an approximation to localized chucking displacements of the wafer coming due to localized shape variations. (b)The IPD maps are dimensionalised using plane-strain approximation [9]. (c) The IPD map is subjected to 6 parameter modelling (translations, magnifications and rotations, in x-y directions each) and the residual map is plotted. (d) The residual map from panel (c) is subjected to a field layout (26mm width, 33mm height) and 6 parameter per-field corrections are modelled. (e) The residual map from (c) is re-modelled on a EUV half-field layout (26 mm wide, 16.5 mm height) on the same 6 parameter terms. (f) The residual map from panel (c) is re-modelled on the advanced packaging scanner field layout (XT260, 52 mm width, 66 mm height), on the same 6 parameter model. In all figures red dashed line marks 147 mm radius, and black dashed line marks 150 mm radius. Data outside 147 mm radius (red dashed line) has been discarded.

strongly indicative that this stress fingerprint originates from unbalanced adhesion forces near the bond front when the bondwave propagation is interrupted by imposing the `final gap' $G$ between the chucks.

### C. *Persistent fingerprint in the shape and the means to control it*

The location of this ring fingerprint is found to vary only with the parameter *final gap*, $G$. This was tested by bonding with different contact pressures, bonder chuck speed, different plasma activation recipes or with a thermal oxide bonding interface instead of SiCN. Thus the occurrence of this shape-discontinuity is due to geometric interference in the natural propagation of bondwave. The results showing the variation with $G$ are plotted in Fig 3(d). Still using the blanket SiCN wafers, parameter $G$ was varied between 20 and 100 microns. A smaller $G$ resulted in this feature appearing closer to the wafer edge. As shown in Fig 3(d), its location varied between 2 and 5 cm from the wafer edge: plotted in the figure as 'distance from the wafer-edge': $R - r_f$, vs $G$. In log-log scale the trend roughly follows the power-law 3/5, which is known to be the natural shape that free-hanging wafers take during bond front propagation in atmospheric conditions [7, 8].

### D. *In-plane-displacements (IPD) using gradient and Stoney model and CPE correction*

The first-derivatives of bonded wafer shapes are computed to interpret the grid distortion in the bonded stack due to sharp changes in wafer geometry. While the shape gradients are visible in raw PWG shapes (Fig 3(a)), the expectation is that these appear as distortion of the litho grid after the wafer to be exposed again is thinned and flat-chucked in a lithography scanner [10,11]. For the bonded wafer shown in Fig 3(a), the full map of in-plane displacements (IPD) is determined by computing local gradients, as shown in Fig 4(a). Naturally an IPD map is non-dimensional, and is dimensionalised using Stoney's approximation for uniform thin-film loading over the surface of a plate [12,13]. Effectively, the non-dimensional displacement fields along direction $i (= x, y)$ are multiplied by some constants $c_i$, which are defined as $c_i = 1/(6 \times T \times E_{2i})$, where $T$ is the thickness of the bonded stack, and $E_{2i}$ is the plane-strain modulus, defined as $E_{2i} = E_i/(1 - \nu_{ij} \times \nu_{ji})$. $E_i$ are the material's Youngs modulus along each $i$ direction, $\nu_{ij}$ are the Poisson's ratio of the anisotropic material, and $\nu_{ji}$ modified as $\nu_{ji} = \nu_{ij} * E_j/E_i$. Here for simplicity, we have ignored the presence of ~200 nm thick bonding interface due to 1550 $\mu$m thick Si(100) crystalline material, and only the wafers' elastic properties have been used. The dimensionalised IPD map is shown in Fig 4(b). These IPD maps are then modelled on a simplified wafer alignment 6-parameter set of equations:

$$dX_i = T_i X_i + M_i X_i + R_i X_j, \quad (2)$$

where the indices (*i,j*) denote (*x,y*), $X$ the coordinates, $T$ the translation terms, $M$ the scaling terms and $R$ (noticeably the cross term), the rotation terms. The residuals after modelling on the map in Fig 4(b) are shown in Fig 4(c).

The dimensionalised IPD maps post 6-parameter alignment were split into typically used field layouts for exposure on ASML scanners. Three layouts in particular are chosen to show the extent of residual distortion per field, post a very basic 6par correction: 26 mm × 33 mm, being the max field size on most twinscan scanners (Fig 4 (d)), 26 mm × 16.5 mm as the 'half-field layout' used in high-NA scanners due to anamorphic optics (residuals shown in Fig 4(e)), and 52 mm × 66 mm layout possible in XT260 scanner, specially targeted for advanced packaging applications (Fig 4(f)). All CPE results in Fig 4 show low levels of distortion coming from bonding despite using low order wafer alignment and CPE models. All these wafers maps contain dense metrology data (~280k points per wafer) and have not been down-sampled to a certain number of markers per field prior to applying the correction models. As expected from previous sections, the ring region is highlighted as containing fields with high distortion, while much of the remaining fields on the wafers can be expected to be easily corrected. Next we turn to using patterned wafers with Cu-filled overlay and alignment marks, and confirm the distortion patterns and trends seen with PWG5-gradient-Stoney model in this section.

## IV. Bonds using Patterned Wafers – Aligned Bonding and Dense Readouts

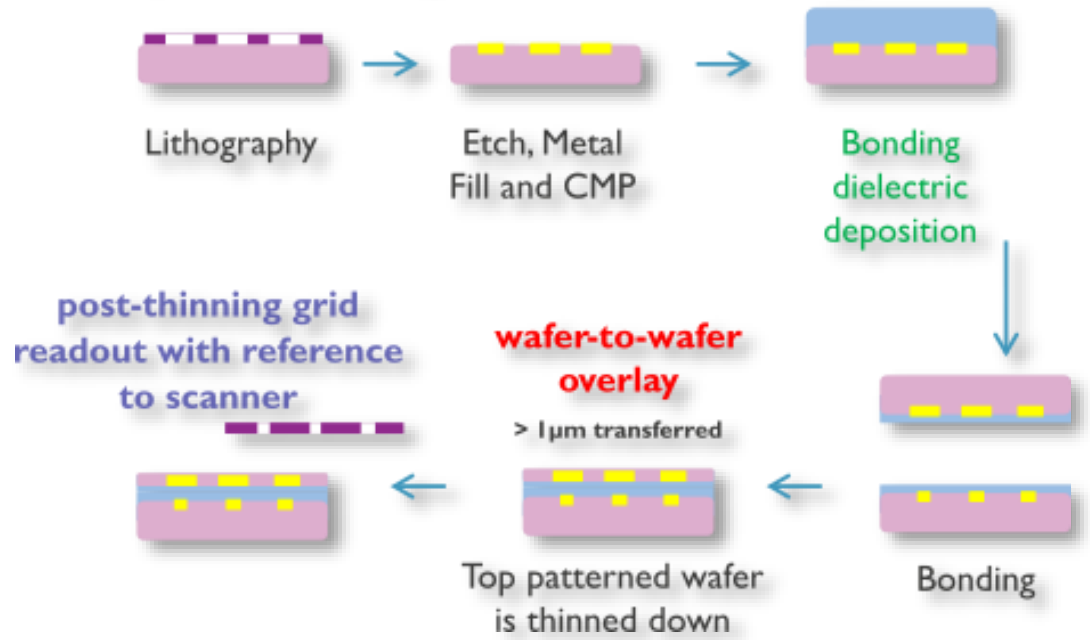

*Fig. 5* Schematic of preparation of wafer pairs with metal markers (etched in $SiO_2$, not shown in the schematic), bonding, and thinning to enable dense readouts of the thinned wafer. Comparing the read-out grid post thinning to previously known grid of the top wafer from marker creation allows us to determine the distortion induced by the process and assembly steps in between marker preparation and back-side litho.

The stack in the patterned wafers is constructed by sequentially depositing dielectric layers on a silicon substrate. The stack consists of Cu alignment and overlay markers made using copper damascene method, etched in 150 nm deep $SiO_2$ layer, and finished with plating, CMP, and further depositions including 30 nm thick SiCN layers, 250 nm $SiO_2$ and finally ~25 nm thick SiCN as the bonding interface. Both the device and carrier wafers are made with the similar stack, except for the patterns etched into each pair to make the markers (which are complementary pairs).

### A. Enabling dense scanner metrology

The bonding of patterned wafer pairs is done using the same hydrophilic bonding sequence as discussed above in sections II and III. A schematic in Fig 5 shows very briefly the flow of using these patterned wafers to bond, thinning of the top wafer to 5 $\mu$m, use of an ASML NXT:2000i for dense readouts of the overlay markers with respect to the scanner grid. After the thinning, both the top (thinned wafer), and the bottom wafer markers are also measured. The measurements are done at 54 points-per-field (ppf), resulting in approximately 13000 markers per wafer (including the partial fields). A direct subtraction of the layers gives the bonding induced overlay between the two wafers, which is frozen at the interface and does not change during thinning (shown in Fig 6 after a 4-parameter removal with translations in *x-y*, uniform rotation and scaling terms).

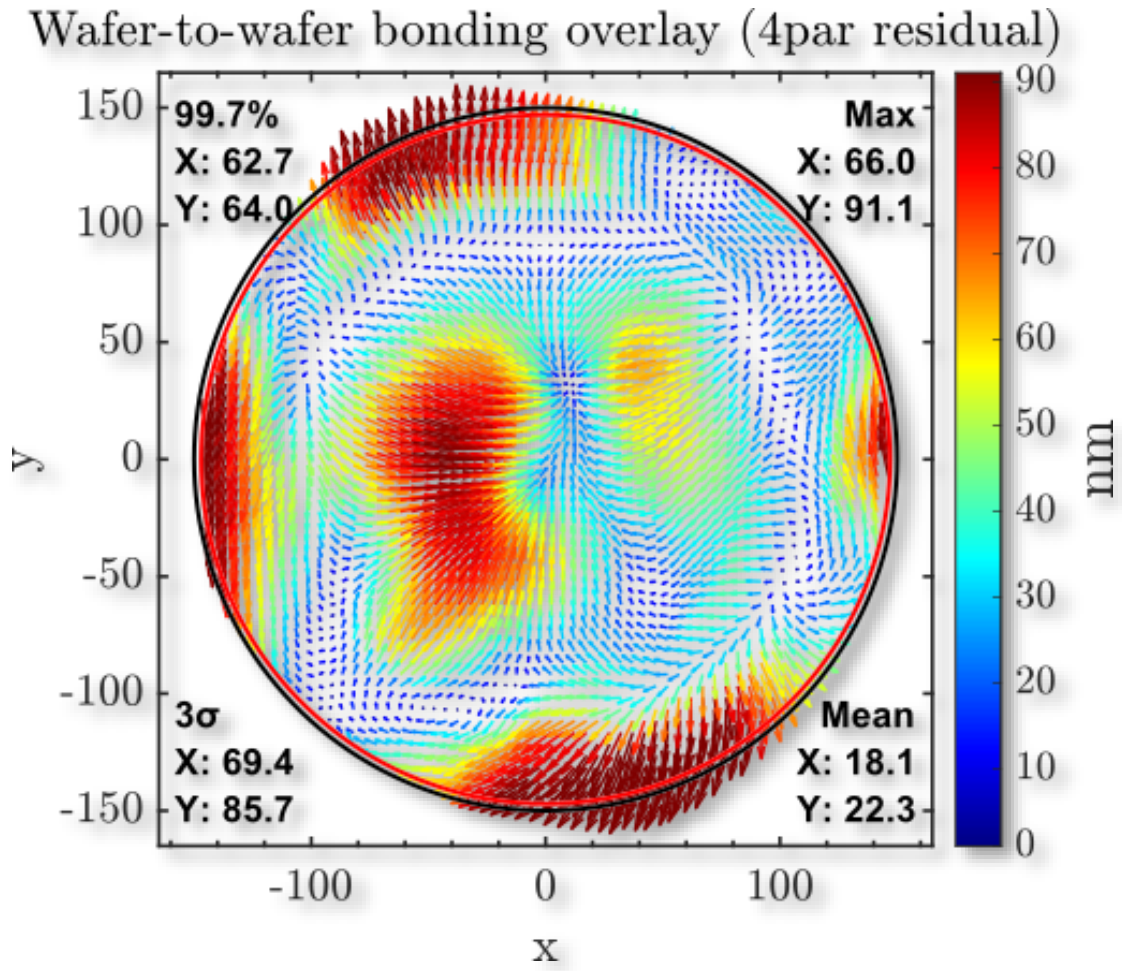

*Fig. 6* Grid measurements of the top and bottom wafer with respect to the scanner grid can be directly subtracted to reveal the misalignments between the bottom-top wafer frozen at the bonding interface during bonding. The above map shows bonding overlay after 4-parameter removal. Data outside 147 mm radius has been discarded.

### B. Modelling the extent of distortion expected on the device wafer post-thinning

The dense scanner readouts with respect to the scanner grid, are then used to model CPE corrections with the field layout of the used wafers. The results are shown in Figure 7. Figure 7(a) shows the wafer map post a basic 6-par alignment model applied. Figure 7(b)—(f) show successively higher order CPE models (6 to 38 par) modelled on the residuals post a 6-par wafer alignment map of Figure 7(a). As found with blanket wafers, the high distortion region is more prominent in a ring shaped region, occurring at the same location when using $G = 30\,\mu$m at bonding. Large vectors at the edge occur typically due to etch [14] and partly due to CPE modelling settings.

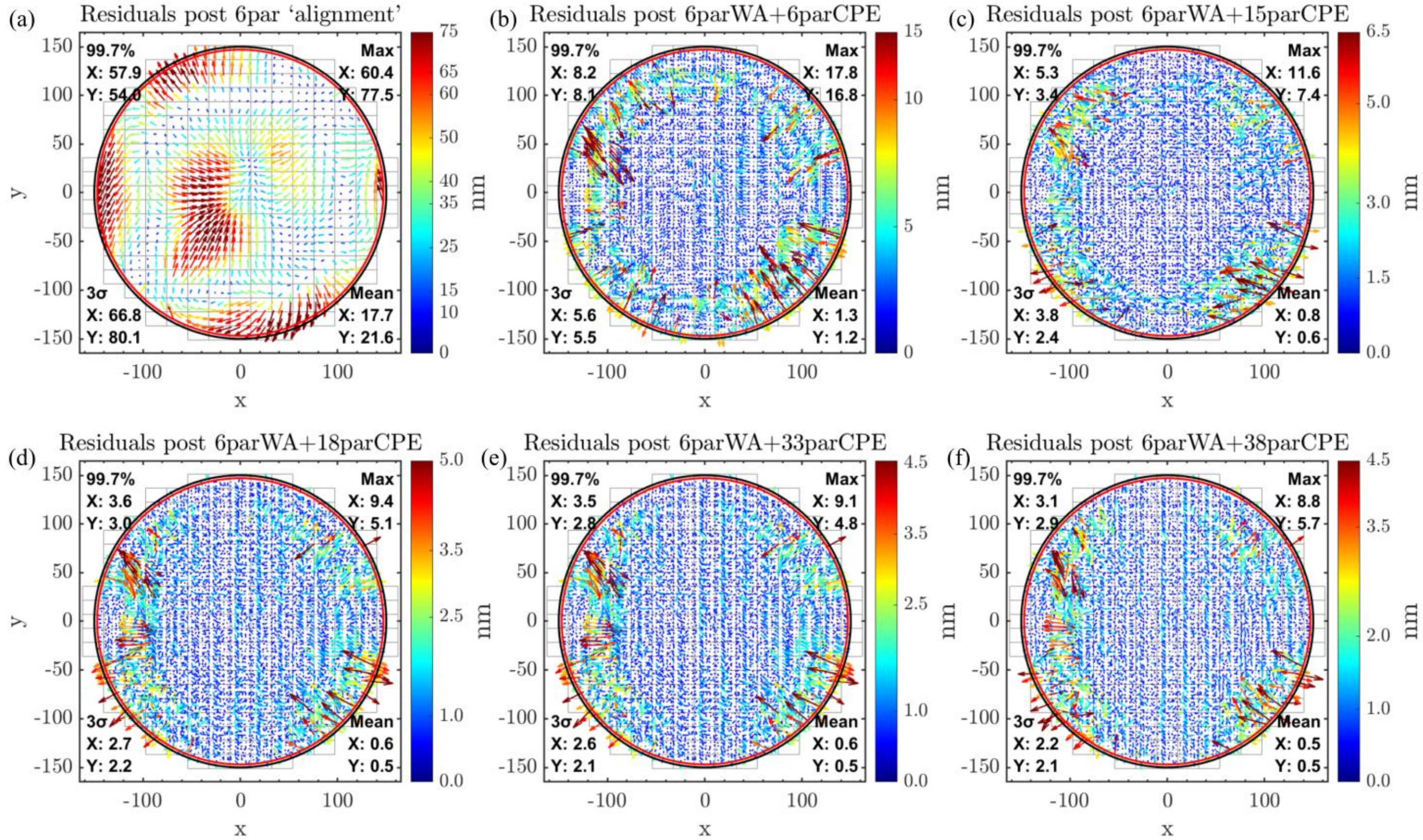

*Fig. 7* Scanner readout of the grid of the top wafer in typical bonded pair with alignment markers on both wafers, measured in ASML NXT:2000i after thinning. The readouts are done at 54 ppf (~13000 per wafer), but the plots show down-sampled distributions (upto 6000 marks per wafer) for clarity. (a) Shows the residuals after aligning the top wafer's litho grid on the simplest 6 parameter model (wafer align, 'WA'). Panels (b—f) show successively higher order (6, 15, 18, 33, 38) correction models applied to the aligned grid in panel (a). The grid layout used here reflects the field layout used at litho step to produce these patterned wafers (21.6 mm width, 14.4 mm height). The circular region persistently show high order distortions, similar to the PWG-derived (IPD) distortion maps in Fig 4. Data outside 147 mm radius (red circle) has been discarded.

### *C. Controlling the high distortion region on patterned wafers*

As noted by using blanket wafers, the nanotopography and in-plane stress created by the interruption of bondwave propagation creates the ring fingerprint. It's a function of adhesive properties of the bonding interface, and thus depends on the surface preparation itself. It would occur in any bonding technique or configuration if there is an interruption to the bond front's natural speed of propagation. It was also identified that the high distortion region's location can be controlled during bonding – by varying the amount of 'final gap' where the chucks stop moving towards each other prior to the vacuum release at wafers' edge. However, the amount of distortion it creates for ultimately performing further lithography steps on one of the wafers, depends on the target matching and the correction capabilities of a scanner.

Here we combined both these insights to use scanner metrology with patterned wafers to show how to minimize the residual grid distortion by using the simple parameter *G*. Several patterned wafer bonds were done while varying *G* between 10 and 100 microns. The top wafer material was removed until 5 microns was left, as indicated in Fig 5. Dense scanner grid readouts (example Figure 8(a,b)) and PWG were used to identify the location of the high-distortion ring region. These are plotted in Figure 8(c) (with same definitions of the ring location ($R - r_f$) and *G*, as Fig 3(d)).

As expected from the results in section II, the ring region moves inwards with increasing *G*, with the high distortion region it creates. For the smaller *G* probed, the ring region lies close to the wafer edge, and thus interacts with high distortion that is created due to natural bonding dynamics (acceleration of the bond front, abruptly limited wafer material at the edge) which is known in the community as a difficult region to control, to obtain more reproducible behavior and higher yielding dies.

The grid distortion residuals from dense scanner readouts while varying G during bonding of the patterned wafers are summarized in Fig 9, using the common lithography overlay statistical metrics over the whole wafer grid: 99.7 percentile of measured overlay magnitudes, mean+3×standard deviation, and the maximum overlay vector. As expected, the higher order models leave lower residuals post the (modelled) correction. Some exceptions occur where 38par model leaves large residuals than 33par, but these are more indicative of correctability by NXT immersion (supporting up to 38 par corr-ection models) or NXE EUV tools (supporting up to 33par models). Having the high distortion region closer to the wafer edge leaves higher magnitudes of residual grid distortion – as it interacts with other sources of the grid distortion (the uncontrolled bonding dynamics and higher variability process steps such as thinning/etch/CMP steps) at the edge.

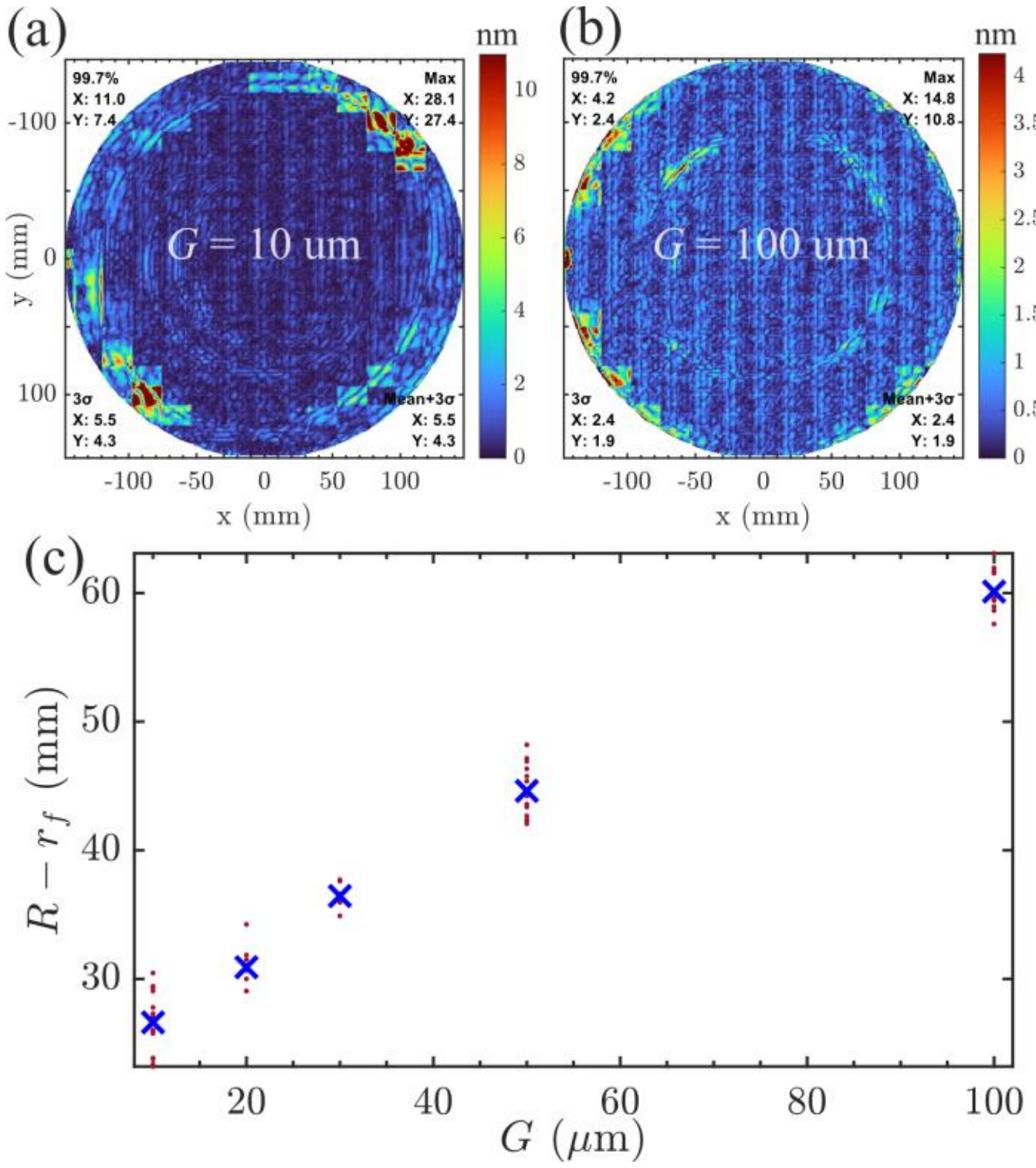

*Fig. 8* Panels (a—b) above show the residual maps of top wafer's grid, with reference to scanner grid, post 6 parameter wafer alignment modelling, and 38 parameter CPE modelling. Two maps ($G = 10, 100$ $\mu$m) are shown, which are the first and last points in panel (c). (c) The location of high-distortion ring region in bonded wafers with markers was identified with scanner readouts and is plotted as a function of the parameter $G$ (final gap, see figures 1 and 3). Placing the wafer chucks further apart interrupts the bond wave farther from the edge, similar to Fig 3(d).

## V. Conclusions

In conclusion, we have shown using blanket-blanket and patterned wafer bonds that the technique used in SUSS XBA by warping the chucked wafers using a small gas pressure from the chuck, naturally yields low distortion bonds. All bonds were done with SiCN as the bonding interface, and prepared using CMP, and standard hydrophilic bonding conditions (plasma activation and DI water rinses).

Using PWG measurements, we revealed the spatially slow and rapidly varying fingerprints in the bonded stack. The technique of initiating the bond here, by bringing into contact wafers bowed towards each other with a gas pressure on the back side, leaves no chuck fingerprints. However a persistent, symmetric ring fingerprint was found in the bonded wafer shapes, whose origin was explained using nanotopography and in-plane stress calculations as a residue to interruption to the bond front.

This region appears as a high distortion region in the lithography grid, once the target wafer is thinned to be exposed in a lithography scanner. The means to controlling the location of this high distortion region in the XBA was identified – by changing the 'final gap' – the distance at which the wafer chucks stop moving towards each other while driving the bonding. The Stoney and 'gradient-model' were used to estimate litho grid distortion from IPD maps (also from PWG measurements). Along these lines, CPE modelling was also done with different field layouts (standard ASML maximum twinscan reticle size, half field layout and the XT260 field layout) to reveal that bonding-induced distortions from this technique are low over most of the wafer.

The distortion determination with blanket SiCN wafers was repeated with patterned wafers containing Cu filled alignment and overlay markers. This enabled scanner metrology, the definitive benchmark for determining how the scanner will see the incoming grid distortion for an application such as back-side power delivery network requiring a back-side print after thinning of one of the wafers in the bonded stack.

Dense scanner metrology (54 ppf, ~13000 markers per wafer) confirmed the distribution and magnitudes of grid distortion estimated from PWG measurements. CPE modelling from 6par to 38par was done on the grid measurements on the thinned wafer, with a field layout representative of the maskset used. The residual grid distortion was summarized in Fig 9. Sub-10 nm residuals in the 99.7%ile metric were found using just 6par CPE modelling. While 38par model pushed it further down to ~3 nm.

The high distortion region was moved to different radial locations on the wafer by tuning the final gap $G$ parameter. It was shown that if this bondwave-interruption occurs closer to region with higher variability due to other sources (here, the edge) the litho correction is less effective in compensating for the cumulative distortion. We conclude that while the origin for this ring fingerprint distortion region is physical (the mechanics of bond front propagation), its interaction with other physical sources of distortion (such as bondwave acceleration, edge voids, process artifacts etc.) must be minimized to achieve high yielding back-side print overlay near the edge.

## Acknowledgment

This work has been enabled in part by the NanoIC pilot line. The acquisition and operation are jointly funded by the Chips Joint Undertaking, through the European Union's Digital Europe (101183266) and Horizon Europe programs (101183277), as well as by the participating states Belgium (Flanders), France, Germany, Finland, Ireland and Romania. For more information, visit nanoic-project.eu.

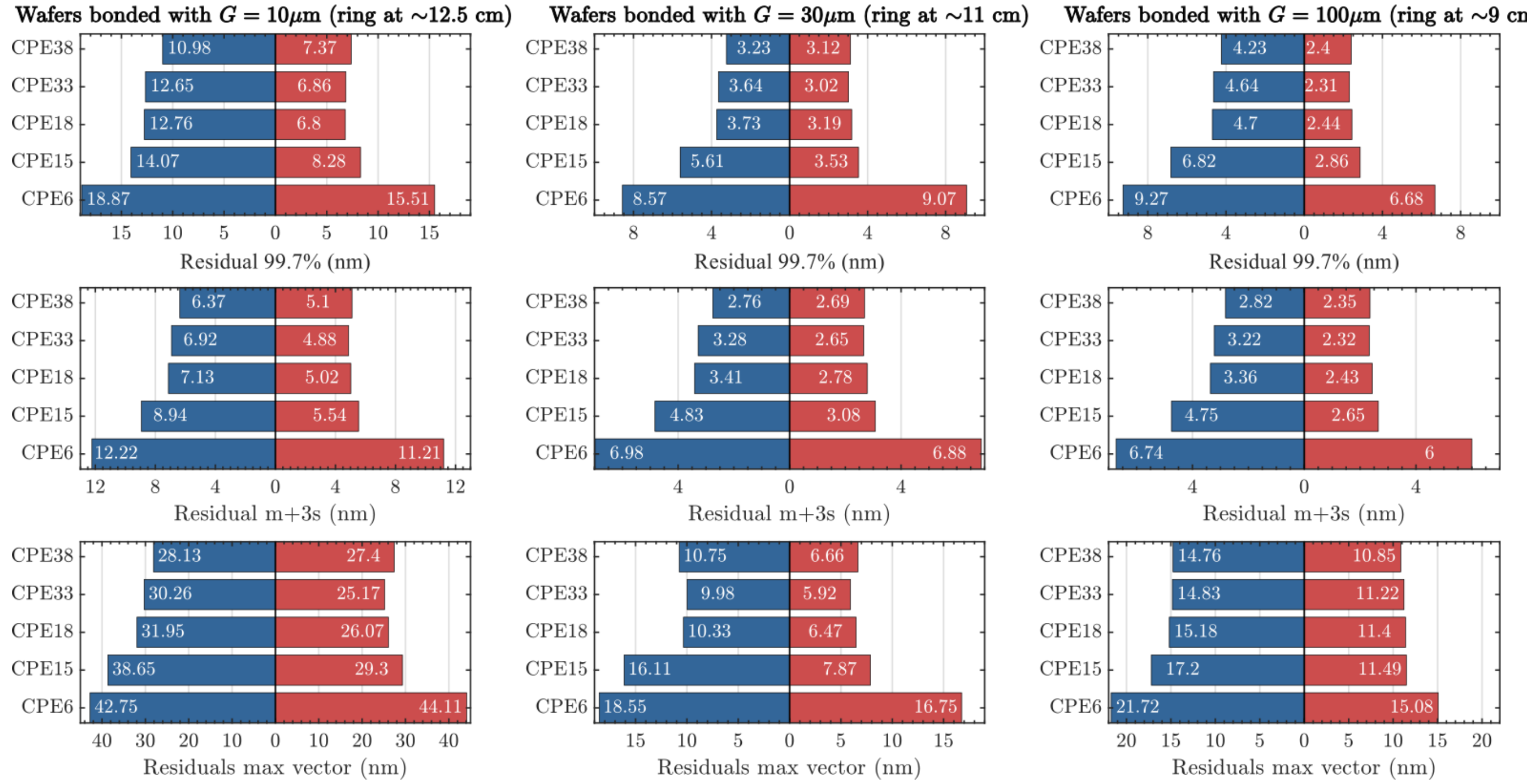

*Fig. 9* Residuals calculated by modelling the different correction models (6—38 par) on the field layout (field width 21.6 mm, height 14.4 mm) shown in Fig 7 and 8 on the thinned wafers' grid readouts from scanner. Three metrics are used for all: 99.7 percentile, mean+3$\sigma$ and the max vector size of the residuals. Three wafers are shown, each of which were bonded while stopping the bondwave (by means of stopping the bonder chuck) at different distances from the center (~12.5, ~11 and ~9 cm). As noted in previous figures, the persistently high distortion region lies distributed a ring region at the mentioned location. The location of this high distortion region was determined by parameter $G$. In general, albeit with exceptions, higher order correction models appear to leave lower backside print overlay.